\documentclass[%
preprint,
 amsmath,amssymb,
 prb,
]{revtex4-2}
\usepackage[utf8]{inputenc}
\usepackage{multirow}
\date{November 2022}
\usepackage{graphicx}
\usepackage{bm}

\usepackage[dvipsnames]{xcolor}

\begin{document}

\title{Imaging the Electric Field with X-Ray Diffraction Microscopy}

\author{Trygve Magnus Ræder}
 \affiliation{Department of Physics, Technical University of Denmark.}
\author{Urko Petralanda}
 \affiliation{Department of Physics, Technical University of Denmark.}
\author{Thomas Olsen}
 \affiliation{Department of Physics, Technical University of Denmark.}
\author{Hugh Simons}
 \affiliation{Department of Physics, Technical University of Denmark.}

\date{\today}% It is always \today, today,

\begin{abstract}

The properties of semiconductors and functional dielectrics are defined by their response in electric fields, which may be perturbed by defects and the strain they generate. In this work, we demonstrate how diffraction-based X-ray microscopy techniques may be utilized to image the electric field in insulating crystalline materials. By analysing a prototypical ferro- and piezoelectric material, BaTiO$_{3}$, we identify trends that can guide experimental design towards imaging the electric field using any diffraction-based X-ray microscopy technique. We explain these trends in the context of dark-field X-ray microscopy, but the framework is also valid for Bragg scanning probe X-ray microscopy, Bragg coherent diffraction imaging and Bragg X-ray ptychography. The ability to quantify electric field distributions alongside the defects and strain already accessible via these techniques offers a more comprehensive picture of the often complex structure-property relationships that exist in many insulating and semiconducting materials.

\end{abstract}
\maketitle
\section{Introduction}
The properties and operation of semiconductors, insulators and, in particular, functional dielectrics, can strongly depend on the shape of the electric fields permeating them. Furthermore, the internal electric field distribution within a crystalline solid is influenced by significant intrinsic and extrinsic factors, such as local anisotropy, strain, the distribution of mobile and immobile charge carriers, as well as a plethora of other defects \cite{lines2001principles, pak1990force}. Understanding the origin of the macroscopic properties of these materials and the devices they comprise requires surveying how these aforementioned features interact with the electric field.

Techniques for imaging the electric field have subsequenly been developed that have the potential for uncovering the driving force of electrons in semiconductors and revealing the space charge around defects. This is particularly relevant for ionic conductivity, where the electric field is both a driver and a result of the ionic charge. Additionally, imaging the electric field provides insight into the local dielectric polarization in functional dielectrics, including how polarization interacts with internal grain boundaries, domain walls and other defects. As X-ray and electron diffraction can already image  strain with high accuracy, the additional capability of imaging the electric field potentially allows us to see the entirity of the electromechanical interactions that dominate the properties of many functional dielectrics. This work details the relationship between the electric field and the crystal lattice, and explores how X-ray scattering in particular may serve as a sensitive probe of the electric field.

There are major challenges to all existing approaches for imaging the electric field, however. Transmission electron microscopy \cite{hachtel2018sub,shibata2015imaging} restricts the sample geometry by requiring thin lamella, and the electron flux of the microscope may additionally affect the distribution of an applied electric field. Kelvin probe force microscopy \cite{melitz2011kelvin} can image the electric field, but is limited to open surfaces, which greatly limits the types of samples that can be investigated, particularly in-operando. Consequently, none of the established characterisation methods can provide a comprehensive picture of electric field distributions in bulk materials, which comprise many of the key applications of, for example, insulators and dielectrics.

Imaging with Bragg-diffracted X-rays may provide a new route to mapping the electric field in bulk materials. Several new quantitative diffraction-based imaging techniques have emerged in the past decade, such as  dark-field X-ray microscopy (DF-XRM)\cite{simons2015dark} and X-ray Bragg coherent diffraction imaging (Bragg-CDI)\cite{shi2022applicability}, Bragg scanning probe X-ray microscopy (Bragg-SPM)\cite{chahine2014imaging}, and Bragg X-ray ptychography\cite{pfeiffer2018x}, which now allow us to obtain images of the scattering function inside materials. Typically, this scattering function is used to map the local distribution of strain inside a single crystal or a single grain in a polycrystalline matrix\cite{simons2018long,shi2022applicability}, since the location of the intensity maxima in reciprocal space can be directly related to the strain. However, since the X-ray scattering function also depends on the position of atomic sublattices in the material and thus the electric polarization of the lattice, it is conceivable that DF-XRM, Bragg-CDI, -SPM or -ptychography could image the electric field within bulk materials directly and in-situ.

\begin{figure}[ht]
    \centering
    \includegraphics[width = 0.65\textwidth]{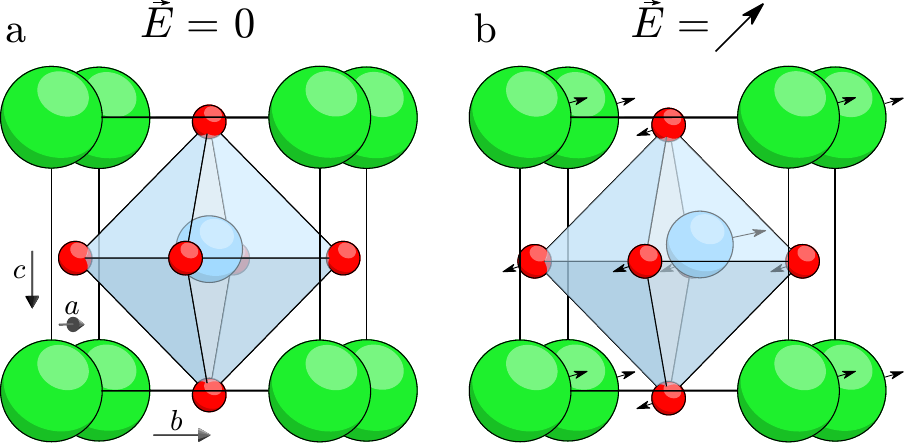}
    \caption{Ionic positions in the BaTiO$_3$ unit cell in \textbf{a} the absence and \textbf{b} presence of an electric field. The motion of each ion is determined by the interatomic potentials, ionic charge, and the direction of the electric field. 
    In \textbf{b} the field is applied equally along the [010] and [001] axes while anisotropy in the interatomic potentials results in larger shifts along the [010] axis than along the [001] (polar) axis. Exaggerated shift magnitudes are shown for illustrative purposes.}
    \label{fig:unit_cell_motion}
\end{figure}

The fact that the scattered intensity is related to the polarization in the material is well known, and has previously been used to study ferroelectric switching in ferroelectric ceramics and films by X-ray diffraction\cite{gorfman2016simultaneous, grigoriev2006nanosecond}. Ionic displacements associated with dielectric polarization will also influence the phase of the scattered x-rays, as demonstrated by  Bragg-ptychography on a PbTiO$_3$film\cite{hruszkewycz2013imaging}. %shabalin2021multiwavelength
%While both phase and intensity of the scattered field was reconstructed in \cite{hruszkewycz2013imaging}, only the phase was used to reconstruct the domain pattern (assuming only up and down domains), while the intensity was disregarded.
%The authors of \cite{hruszkewycz2013imaging} also note variation in the intensity across the sample, and show theoretically that the intensity will depend on the magnitude of polarization within each domain.
%During ferroelectric switching the polarization alternates between bi-stable states and this provides relatively large sublattice shifts compared to electronically induced displacement. 
%The x-ray scattering scattering function is related to the electric field as both concern the motion of ions.
%In fact, there are two different routes to obtaining the electric field inside a material from the scattered field: via the intensity\cite{gorfman2016simultaneous,shabalin2021multiwavelength} or via the phase\cite{shabalin2021multiwavelength}. The motion of the anion and cation sublattice(s) in the presence of an electric field is illustrated in Figure \ref{fig:unit_cell_motion} for the case of BaTiO$_3$.
%X-ray diffraction is a technique that examins the average structure of the mateiral. 
Based upon this prior work, it is conceivable to image the electric field with the use of X-ray diffraction microscopy to obtain spatial information about the distribution of the electric field. As the electric field drives a smaller sublattice shift than that associated with ferroelectric switching, this would require a low noise level in the instrumentation to provide quantitative data.
Additionally, a homogeneous strain state was assumed in \cite{hruszkewycz2013imaging}, and for a general case where both the polarization and strain varies, the strain and polarization information will be convoluted in the phase of the scattered X-ray beam. Shablin et al.\cite{shabalin2021multiwavelength} suggested a method for separating the strain and polarization information contained in the phase of the scattered beam, by choosing two X-ray energies on either side of an absorption edge. However, if instead the electric field is extracted via the intensity of the diffracted beam, strain and polarization will be encoded in different aspects of the beam. The strain field will be encoded in the scattering direction, while the polarization will be encoded in the total scattering intensity. This way, a single wavelength may be used to obtain both strain and polarization even in non-ideal samples with complicated strain fields.

In this work, we explore the potential of imaging the electric field in X-ray microscopy utilizing either the intensity or phase. We first quantify how atomic displacements affect the phase and intensity of the scattering function, and then quantify the atomic displacements in the unit cell brought on by an electric field. By combining the above, we obtain relationships between the electric field and scattering phase and intensity. These relationships are then interpreted as scaling laws that determine the signal-to-noise ratio to be overcome in order to recover the electric field from experimental data. Finally, we will discuss how the obtained scaling laws can guide experimental design, and use these to propose an experimental configuration to realize the approach.

%In reality, the dielectric constant depends on the applied electric field, and a more accurate Equation~would use Landau-Devonshire-Ginzburg (LDG) theory to calculate $\frac{dP}{dE}$ as a continuous function on $E$.

%(300-1)*8.854*10**-12((4*10**-10)**3)/(4*1.602*10**-19)*(10**5)
% => 2.6440409e-14 m
%C/V/m * mmm * 1/C * V/m = m 
%1 kV/cm = 1000*10**2 = 10**5 V/m 
\section{Approaches for polarization retrieval}

Sensitivity to the electric field, either via intensity or phase, will arise from the scattering function $F_{hkl}$. 
By assuming a linear relationship for relevant fields, the impact of an electric field on the intensity $|F_{hkl}|^2$ and the phase $\phi_{F_{hkl}}$ of the scattered field may be expressed as follows: 
\begin{align}
    \label{eq:df2dE0}
    |F_{hkl}(\textbf{E})|^2-|F_{hkl}(0)|^2 =&\  \nabla_{\textbf{E}} |F_{hkl}|^2\cdot\textbf{E}\\
    \label{eq:dphidE0}
    \phi_{F_{hkl}}(\textbf{E})-\phi_{F_{hkl}}(0) =&\ \nabla_{\textbf{E}} \phi_{F_{hkl}} \cdot\textbf{E}
\end{align}
Using Equations \ref{eq:df2dE0} and \ref{eq:dphidE0}, experimentally obtained intensity or phase ($|F_{hkl}|^2$ or $\phi_{F_{hkl}}$) can be used with theoretically obtained sensitivity ($\nabla_{\textbf{E}} |F_{hkl}|^2$ or $\nabla_{\textbf{E}} \phi_{F_{hkl}}$) to map the electric field. 
This section describes how the intensity and phase may be obtained experimentally, while the next section details how the sensitivity may be obtained.

\subsection{The intensity $|F_{hkl}|^2$}
The displacement field may be found via the intensity only if the amplitude of the structure factor squared, $|F_{hkl}|^2$, is known with high precision at each position in the sample. For a homogeneously strained state, the entire sample scatters at the same diffraction condition, and  $|F_{hkl}|^2$ may be mapped from a single real-space image. However, in the general case of an arbitrary strain state different regions of the sample scatter in (slightly) different directions, and an integration over reciprocal space is required. We note that many relevant materials are piezoelectric, and therefore generate an inhomogeneous strain field in the presence of an inhomogeneous electric field. The experimental implementation will be different depending on the microscopy technique used, but in general the three dimensions of reciprocal space near a reflection $Q$ represent three angles, two representing orientation of the sample (i.e. $\phi$ and $\chi$ in DF-XRM) and the scattering angle $2\theta$. However, a convergence in the incoming beam, or an energy spread will allow a larger part of reciprocal space to be probed simultaneously.
%Experimentally, this requires a three dimensional series of images:
%\begin{equation}
%    |F_{hkl}|^2 = \int_{\phi}\int_{\chi}\int_{2\theta} I 
%\end{equation}
%The total intensity $|F_{hkl}|^2$ will relate to atomic displacements as each atom scatters with a different amplitude and phase. The scattering phase of each atom is also affected by anomalous scattering, which has a different impact on $hkl$ and $\bar{h}\bar{k}\bar{l}$ reflections. 

\subsection{The phase, $\phi_{F_{hkl}}$}
In addition to intensity, the phase of the scattered field can readily be obtained by a variety of methods, including ptychographic DF-XRM\cite{mads_phase_recovery}, Bragg-CDI or Bragg-ptychography. 
The phase contains information about the displacement of the unit cell in relation to its neighbours ($\phi_\text{disp}$) and displacement of individual atoms in the unit cell (polarization $\phi_\text{pol}$), as follows:
\begin{equation}
    \phi_{F_{hkl}} = \phi_\text{disp} + \phi_\text{pol}
\end{equation}
Here $\phi_\text{disp} = \textbf{u}\textbf{Q}$ where $\textbf{u}$ is the displacement field and $\textbf{Q}$ is the scattering vector. 
%The spatial derivative of $\phi_\text{disp}$ will be proportional to strain, although describing strain components determined by the direction of $Q$ and the oblique imaging plane.
%The two spatial derivatives of $\phi_\text{disp}$ in an image describe one component of shear strain, and one component of mixed shear-normal type determined by the oblique angle.
$\phi_\text{pol}$ describes changes to the structure factor $F_{hkl}$ due to the offset of individual atoms. As realistic atomic displacements are small compared to the unit cell size, $\phi_\text{pol}$ can be assumed linear with the electric field. This gives:
\begin{equation}
    \phi_{F_{hkl}} = \textbf{u}\textbf{Q}  + \nabla_\textbf{E}\phi_{F_{hkl}} \textbf{E}
\end{equation}
The above equation is a linear equation with two unknowns %namely the magnitude of \textbf{u} along \textbf{Q} and the magnitude of \textbf{E} along $\nabla_\textbf{E}\phi_{F_{hkl}}$ 
and may be solved using linear algebra if another complimentary equation is found.
In \cite{shabalin2021multiwavelength} the authors suggest using two X-ray energies on either side of an absorption edge.
%This changes the scattering conditions, but has a negligible impact on $2\theta$.
Other approaches for finding two independent relations include using multiple different (hkl) reflections, or possibly using a thermodynamical potential. %As the specific approach is not the main topic in this publication we have moved a further description of the different approaches to the supporting information.

\section{Experimental sensitivity to the electric field}

The sensitivity of the intensity and phase to the electric field may be expressed as the sum of partial derivatives with respect to ionic position $\textbf{R}_j$ using the multivariate form of the chain rule:
\begin{align}
    \label{eq:df2dE}
    \nabla_{\textbf{E}} |F_{hkl}|^2 =&\ \sum_j \nabla_{\textbf{R}_j} |F_{hkl}|^2  \nabla_{\textbf{E}} \textbf{R}_j\\
    \label{eq:dphidE}
    \nabla_{\textbf{E}} \phi_{F_{hkl}} =&\ \sum_j \nabla_{\textbf{R}_j} \phi_{F_{hkl}}  \nabla_{\textbf{E}} \textbf{R}_j
\end{align}
\subsection{Sensitivity to ionic motion, $\nabla_{\textbf{R}_j} |F_{hkl}|^2$ and $\nabla_{\textbf{R}_j} \phi_{F_{hkl}}$}
We here use BaTiO$_3$ as an example, as it is a prototypical ferroelectric material as well as the most widely used capacitor material.
%, the unit cell contains 102 electrons, with more than half (54) centered on the Ba$^{2+}$ ion, 18 on the Ti$^{4+}$ ion, 10 on each O$^{2-}$. 
In the ground state, the Ti-ion resides 0.18~Å (4.5~\%) off-center, with additional or lesser offset depending on the electric field. The oxygen ions are offset in the opposite direction.
Additionally, different ions scatter x-rays with different phase shifts due to anomalous scattering, an effect that is particularly strong when the X-ray energy is right above an absorption edge. For the particular example of BaTiO$_3$, the absorption edges  are not available because transmission is too low at the Ti K- and Ba L1-edge (5~keV and 6~keV) while scattering is too weak at the Ba K-edge (37.4~keV).
The structure factor $F_{hkl}$ is defined as follows:
\begin{align}
    F_{hkl} = \sum_{j} f_j \mathrm{exp}(-i\textbf{Q}_{hkl} \textbf{R}_j)
    %F = q_{Ba} + q_{O1}e^{i\textbf{Q}\textbf{p}_{O1}} + q_{O1}e^{i\textbf{Q}\textbf{p}_{O1}} + %q_{O2}e^{i\textbf{Q}\textbf{p}_{O2}} +q_{O3}e^{i\textbf{Q}\textbf{p}_{O3}}
    \label{eq:scattering}
\end{align}
%56+22+8*3

\begin{figure}[ht]
    \centering
    \includegraphics[width = 0.75\textwidth]{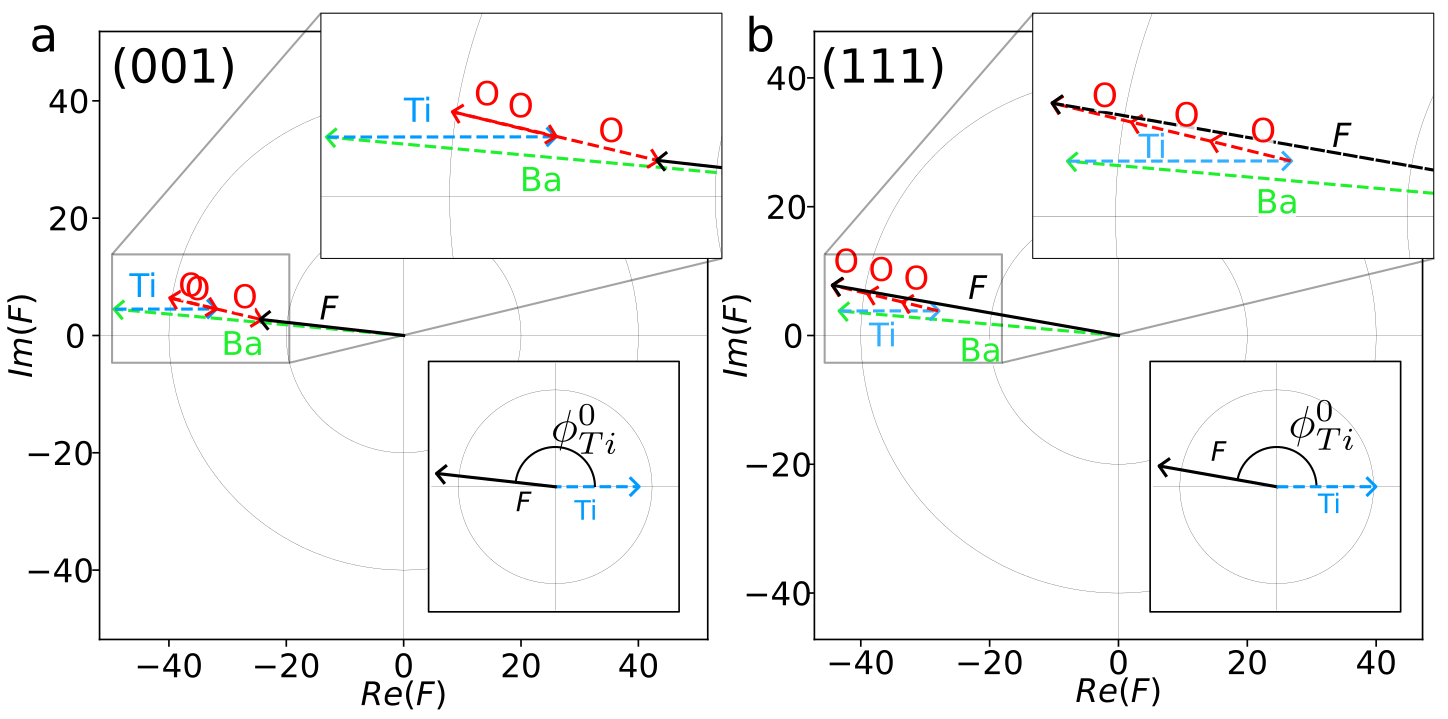} 
    \caption{The structure factors of \textbf{a} (001) and \textbf{b} (111) BaTiO$_3$ at 17~keV, displayed as the sum of scattering from the different sub-lattices and rotated so that scattering from Ti has no imaginary component in the ground state. Each sub-lattice is represented with a dashed line colored according to the respective ion. $\phi_{Ti}^0$ is shown in the insets of the two figures. $\text{sin}(-\phi_{Ti}^0)$ is $\sim$0.10 in \textbf{a} and $\sim$0.17 in \textbf{b}. Consequently the intensity of the (111) reflection has higher sensitivity to the electric field than the (001) reflection.}
    \label{fig:F}
\end{figure}

% motivate
% multiscale problem
% disordered materials
%From Equation~\ref{eq:scattering} it can be seen that the change to $F_{hkl}$ when a single atom moves is proportional to $Q_{hkl}$, i.e. higher hklgive a larger change to $F_{hkl}$.
To derive $\nabla_{\textbf{R}_j} |F_{hkl}|^2$, we offset the $F_{hkl}$ in the real-imaginary plane so that a change in $\textbf{R}_j$ from the ground-state position only affects Im$(F_{hkl})$, as illustrated in Figure \ref{fig:F}. 
We define the offset as $\omega_j^0 = - \text{ang}(f_j)+\textbf{Q}_{hkl} \textbf{R}_j^0$.
We further define $F_{hkl}^0$ as the structure factor, and $\textbf{R}_j^0$ as the position of ion $j$ in the absence of an electric field. The scattering intensity as a function of $\textbf{R}_j$ may be expressed as follows for small displacements:
\begin{align}
    |F_{hkl}|^2(\textbf{R}_j) =&\ |\text{exp}(i\omega_j^0)F_{hkl}|^2 \\
                =&\ \text{Re}\Big\{\text{exp}(i\omega_j^0)F_{hkl}^0 + \text{exp}(i\omega_j^0)(\textbf{R}_j-\textbf{R}_j^0) \cdot\nabla_{\textbf{R}_j}F_{hkl} \Big\}^2
                \nonumber\\
                \label{eq:unportial}
                +&\ \text{Im}\Big\{\text{exp}(i\omega_j^0)F_{hkl}^0 + \text{exp}(i\omega_j^0)(\textbf{R}_j-\textbf{R}_j^0)\cdot\nabla_{\textbf{R}_j}F_{hkl} \Big\}^2
\end{align}
We introduce $\phi_j^0$ and expand the partial derivative to simplify the expression:
\begin{align}
    \phi_j^0 =& \omega_j^0+\text{ang}(F_{hkl}^0) \\
    =& -\text{ang}(f_{j})+\textbf{Q}_{hkl} \textbf{R}_j^0 +\text{ang}(F_{hkl}^0)\\
    \text{exp}(i\omega_j^0)\nabla_{\textbf{R}_j}F_{hkl} =& -\text{exp}[-i\text{ang}(f_j)+i\textbf{Q}_{hkl}\textbf{R}_j^0]i\textbf{Q}_{hkl}f_j\text{exp}-i\textbf{Q}_{hkl}\textbf{R}_j)\\
    =& -i\textbf{Q}_{hkl}|f_j|\text{exp}(-i\textbf{Q}_{hkl}(\textbf{R}_j-\textbf{R}_j^0))\\
    \label{eq:partial_der}
    \approx & -i\textbf{Q}_{hkl}|f_j|
\end{align}
Where Equation \ref{eq:partial_der} has been simplified for small displacements. Equation \ref{eq:partial_der} has no real part, and we find the following by inserting Equation \ref{eq:partial_der} into Equation \ref{eq:unportial} and simplifying with $\phi_j^0$:
\begin{align}
    |F_{hkl}|^2(\textbf{R}_j) 
                =&\ \text{Re}\Big\{\text{exp}(i\phi_j^0)|F_{hkl}^0| \Big\}^2
                \nonumber\\
                +&\ \text{Im}\Big\{\text{exp}(i\phi_j^0)|F_{hkl}^0|  -i|f_j|\textbf{Q}_{hkl}(\textbf{R}_j-\textbf{R}_j^0)\Big\}^2\\
                =&\ \text{cos}^2(\phi_j^0)|F_{hkl}^0|^2 + \Big(\text{sin}(\phi_j^0)|F_{hkl}^0|  -|f_j|\textbf{Q}_{hkl}(\textbf{R}_j-\textbf{R}_j^0)\Big)^2\\
                =&\ |F_{hkl}^0|^2 -2\text{sin}(\phi_j^0)|F_{hkl}^0||f_j|\textbf{Q}_{hkl} (\textbf{R}_j-\textbf{R}_j^0) + (|f_j|\textbf{Q}_{hkl}(\textbf{R}_j-\textbf{R}_j^0))^2
\end{align}
For small displacements we thus have:
\begin{align}
    \nabla_{\textbf{R}_j}|F_{hkl}|^2 = &\
    -2\text{sin}(\phi_j^0)|F_{hkl}^0||f_j|\textbf{Q}_{hkl}
    \label{eq:df2dr}
\end{align}

For the phase we consider only the relative phase  with respect to the phase of the ground state $F_{hkl}^0$. For small displacements we have the following, 
where we have simplified tan$(x)\approx x$ for a small $x$ and substituted $\textbf{R}_j$ by $\textbf{R}_j^0$ to arrive at a linear relationship between phase and atomic displacement:

\begin{align}
    \phi_{F_{hkl}}-\text{ang}(&F_{hkl}^0)=\\ \text{arctan}\Big(&\frac{
        |f_j|
        %\text{sin}(\text{ang}(f_i)-\text{ang}(F_{hkl}^0)-\textbf{Q}_{hkl}\textbf{R}_j)
        %\text{sin}(-\phi_j^0)
        \text{sin}[\text{ang}(f_{j})-\textbf{Q}_{hkl} \textbf{R}_j -\text{ang}(F_{hkl}^0)]
        }{
        |F_{hkl}|
        }\Big)\nonumber\\
    \nabla_{\textbf{R}_j}\phi_{F_{hkl}} =& 
    -\frac{|f_j|}{|F_{hkl}^0|}
    %\text{cos}(\text{ang}(f_i)-\text{ang}(F_{hkl}^0)-\textbf{Q}_{hkl}\textbf{R}_j^0)\textbf{Q}_{hkl}
    \text{cos}(-\phi_j^0)
    \textbf{Q}_{hkl}
    \label{eq:dphidr}
\end{align}

Equations \ref{eq:df2dr} and \ref{eq:dphidr} show how $\phi_j^0$ is critical for the sensitivity of both intensity and phase to ionic motion ($\nabla_{\textbf{R}_j}|F_{hkl}|^2 \propto -\text{sin}(-\phi_j^0)$ and $\nabla_{\textbf{R}_j}\phi_{F_{hkl}} \propto
\text{cos}(-\phi_j^0)$).
In BaTiO$_3$, the Ti ions have the largest ionic motion and $\phi_{Ti}^0$ is shown in Figure~\ref{fig:F}, in the context of the (001) and (111) structure factors of BaTiO$_3$.

\subsection{Change in phase when multiple ions move}
\begin{figure}[ht]
    \centering
    \includegraphics[width = 0.75\textwidth]{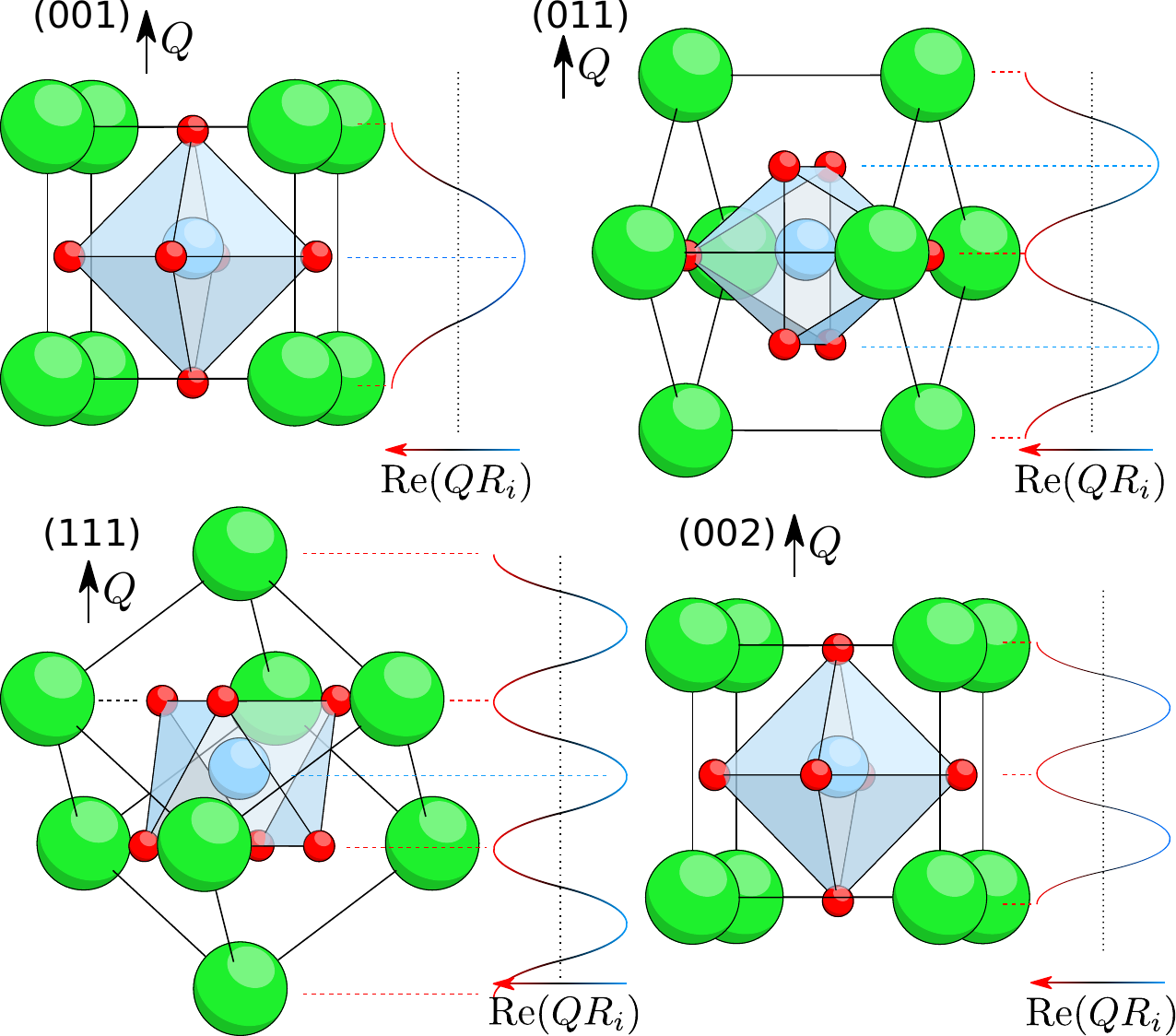}
    \caption{Views of the BaTiO$_3$ unit cell oriented with the (001), (011), (111) and (002) reflections vertical in the figure. The relative phase, $\textbf{QR}_i$, of scattering is indicated next to each orientation. Dotted lines connect ions scatter in \textcolor{red}{positive} and \textcolor{blue}{negative} phase.}
    \label{fig:unit_cell}
\end{figure}

Change in the scattering phase from the motion of multiple ions can be either additive or subtractive. 
The sign  of $\nabla_{\textbf{E}}\textbf{R}_{j}$ depends on the sign of the ionic charge, while $\nabla_{\textbf{R}_j}\phi_{F_{hkl}}$ depends the sign of cos($-\phi_j^0$), i.e. the phase at which the X-rays scatter on the ion.
The scattering phase $\textbf{Q}\textbf{R}_j$ of the different ions of the unit cell is shown in Figure~\ref{fig:unit_cell} for the (001), (011), (111) and (002) reflections.

The number of oxygen ions that scatter out-of-phase with the Ti-ion, are 1, 2 and 3 for the (001), (011) and (111) reflections, respectively, while all ions scatter in-phase for the (002) reflection.
We therefore expect smaller phase shifts for the (001), (011) and (002) reflections compared to the (111) reflection. 

\subsection{Ionic displacements in an electric field, $\nabla_{\textbf{E}} \textbf{R}_j$}

The Born effective charge tensor is defined as the change in polarization due to a small displacement of an atom
\begin{align}
    Z^*_{j,\beta\alpha}=V_{uc}\frac{\partial P_\beta}{\partial R_{j,\alpha}}.
\end{align}
Where we let $V_{uc}$ be the volume of the unit cell. This allows one to define a local polarization associated with a particular atom. Conversely
\begin{align}
    \frac{\partial R_{j,\alpha}}{\partial P_\beta}=&\ V_{uc}Z^{*-1}_{j,\beta\alpha}\\
    \frac{\partial R_{j,\alpha}}{\partial E_\gamma}=&\ V_{uc}\sum_{\beta} \chi_{\beta\gamma}^\text{lattice}Z^{*-1}_{j,\beta\alpha}
    \label{eq:set}
    %\partial R_{j,\alpha}=V\sum_{\beta} Z^{*-1}_{j,\beta\alpha}\partial P_\beta=V\sum_{\beta\gamma} Z^{*-1}_{j,\beta\alpha}\chi_{\beta\gamma}^{lat}\partial E_\gamm
    %\frac{\partial R_j,\alpha}{\partial E_\beta} = &V_uc Z^{-1}_{j,\alpha\alpha}(\varepsilon_0_{\alpha\alpha}-\varepsilon_0^{hf}_{\alpha\alpha})\delta_{\alpha\beta}
\end{align}
where the last equality follows from the definition of the lattice polarizability $\chi_{\alpha\beta}^\text{lattice}$.
The lattice polarizability is related to the dielectric tensor at low frequency, $\varepsilon_{\alpha\beta} = \varepsilon_0\delta_{\alpha\beta}+\chi_{\alpha\beta}^\text{electronic}+\chi_{\alpha\beta}^\text{lattice}$, but not the dielectric constant at high frequency, $\varepsilon^{hf}_{\alpha\beta} = \varepsilon_0\delta_{\alpha\beta}+\chi_{\alpha\beta}^\text{electronic}$.
%$\varepsilon_{\alpha\beta}=\varepsilon_0\delta_{\alpha\beta}+\chi_{\alpha\beta}^\text{lattice}+\chi_{\alpha\beta}^\text{electronic}=\varepsilon_{\alpha\beta}^{hf}+\chi_{\alpha\beta}^\text{lattice}$. 
We thus get
\begin{align}
    \frac{\partial R_{j,\alpha}}{\partial E_\gamma}=V_{uc}\sum_{\beta}(\varepsilon_{\beta\gamma}-\varepsilon^{hf}_{\beta\gamma})Z^{*-1}_{j,\beta\alpha}
    \label{eq:set2}
\end{align}
For BaTiO$_3$ the crystal field of each ion has the symmetry of the unit cell (tetragonal, P4mm) and the three axes are independent, which implies that $\varepsilon_{\alpha\neq\beta}~=~Z^*_{j,\alpha\neq\beta}~=~0$. For BaTiO$_3$ we may thus express Equation~\ref{eq:set2} as follows:
\begin{align}
   \frac{dR_{j,\alpha}}{d\alpha}=V_{uc}(\varepsilon_{\alpha\alpha}-\varepsilon^{hf}_{\alpha\alpha})Z^{*-1}_{j,\alpha\alpha}
   %\nabla_\textbf{E}\textbf{R}_j=V_{uc}(\bm{\varepsilon}-\bm{\varepsilon}^{hf})\textbf{Z}^{*-1}_{j}
    \label{eq:drde}
\end{align}
%Where $\bm{\varepsilon}$ is the vector form of the dielectric constant, consisting of the diagonal elements, $\textbf{Z}^{*-1}_{j}$ is the equivalent for the Born effective charge tensor, and the multiplicaiton is element-wise.
for $\alpha = x,y,z$.

Inserting Equations~\ref{eq:df2dr}, \ref{eq:dphidr}, and \ref{eq:drde} into Equations~\ref{eq:df2dE} and \ref{eq:dphidE} gives the total sensitivity of the intensity and phase to the electric field:
\begin{align}
    \label{eq:fin_f2}
    \nabla_{\textbf{E}} |F_{hkl}|^2 =& \sum_j
    -2|F_{hkl}^0||f_j|\text{sin}(\phi_j^0)\textbf{Q}_{hkl}  V_{uc}(\bm{\varepsilon}-\bm{\varepsilon}^{hf})\textbf{Z}^{*-1}_{j}\\
    \label{eq:fin_phi}
    \nabla_{\textbf{E}} \phi_{F_{hkl}} =& \sum_j
    -\frac{|f_j|}{|F_{hkl}^0|}\text{cos}(-\phi_j^0)\textbf{Q}_{hkl}  V_{uc}(\bm{\varepsilon}-\bm{\varepsilon}^{hf})\textbf{Z}^{*-1}_{j}
\end{align}

which comprises the main findings of this publication. In Equation \ref{eq:fin_f2} and \ref{eq:fin_phi} the atomic form factor may be found from an X-ray database (such as DABAX\cite{dabaxdynamic}), the atomic positions may be found from a crystallographic database, and the dielectric constant may be measured experimentally. The Born effective charge is available from simulations, which also provide an alternate dielectric constant.
Equations \ref{eq:fin_f2} and \ref{eq:fin_phi} show a linear effect with the electric field and greater accuracy at high fields may be achieved if a field-dependent dielectric constant ($\varepsilon(E)$) is used.

Table~\ref{tab:dft} shows data obtained for BaTiO$_3$ with density functional theory (DFT) simulations used to determine the  $\textbf{Z}^*_j$ and the ionic contribution to the dielectric constant for BaTiO$_3$. The DFT calculations used ASE \cite{Hjorth_Larsen_2017}, GPAW \cite{Enkovaara_2010} and Phonopy \cite{Togo2015}. The system was restricted to the room-temperature symmetry (tetragonal) and the lattice constants were fixed to experimental values in order to avoid instabilities that arise in the fully relaxed tetragonal structure \cite{Peng2020}. 
%Further details on the DFT calculations are provided in the Supporting Information. 
Note that we choose to use the dielectric constant from DFT so that is consistent with the values used for the Born effective charge. The value used here ( $\bm{\varepsilon} = (7200,7200,640)$), is therefore higher than typical experimentally reported values $\sim(4000,4000,200)$\cite{merz1949electric} at room temperature.
Comparing the (001), (011) and (111) reflections show how the direction of the $\textbf{Q}$ vector matters strongly. Additionally, higher order peaks generally have higher sensitivity as illustrated by comparing the (001) and (002) reflections.

\begin{table}[ht]
\begin{center}
\caption{ The change in phase and intensity in BaTiO$_3$ when subject to a field of  $|\textbf{E}|$~=~1~kV/cm along the $Q$ vector at 17 keV, a typical X-ray energy in DF-XRM that yields good penetration in BaTiO$_3$.}
    \begin{tabular}{c|llllll}
    \label{tab:dft}
& (001) & (011) & (111) & (002) \\
&\ \ (00-1) &\ \ (0-1-1) &\ \ (-1-1-1) &\ \ (00-2) \\ \hline
\multirow{2}{*}{$\Delta \phi_{F_{hkl}}$\ \ \ $\Big[\frac{\text{mrad}}{\text{kV/cm}}\Big]$}
& -5.34& -20.13& -14.36& -2.00\\ 
&\ \ 5.35&\ \ 20.06&\ \ 14.42&\ \ 2.02\\ 
\multirow{2}{*}{$\frac{\Delta |F_{hkl}|^2}{|F_{hkl}^\star|^2}$ $\Big[\frac{\%}{\text{kV/cm}}\Big]$}
& -0.053& 0.134& -0.257& -0.032\\ 
&\ \  0.035&\ \  0.172&\ \  -0.630&\ \  -0.065
\end{tabular}
\end{center}
\end{table}

\section{Results and discussion}
%Equations~\ref{eq:fin_f2} and \ref{eq:fin_phi} and the results in Table \ref{tab:dft} have a number of implications for using X-ray diffraction microscopy to quantify structure-property relationships in functional dielectrics. 
%We here start by defining the conditions where quantitative imaging of the electric field may be performed. 
%We will then describe the implications for imaging the strain field in X-ray microscopy in the prescence of electric fields.\\

\subsection{Proposed experimental realization}
\begin{figure}
    \centering
    \includegraphics[width = 0.78\textwidth]{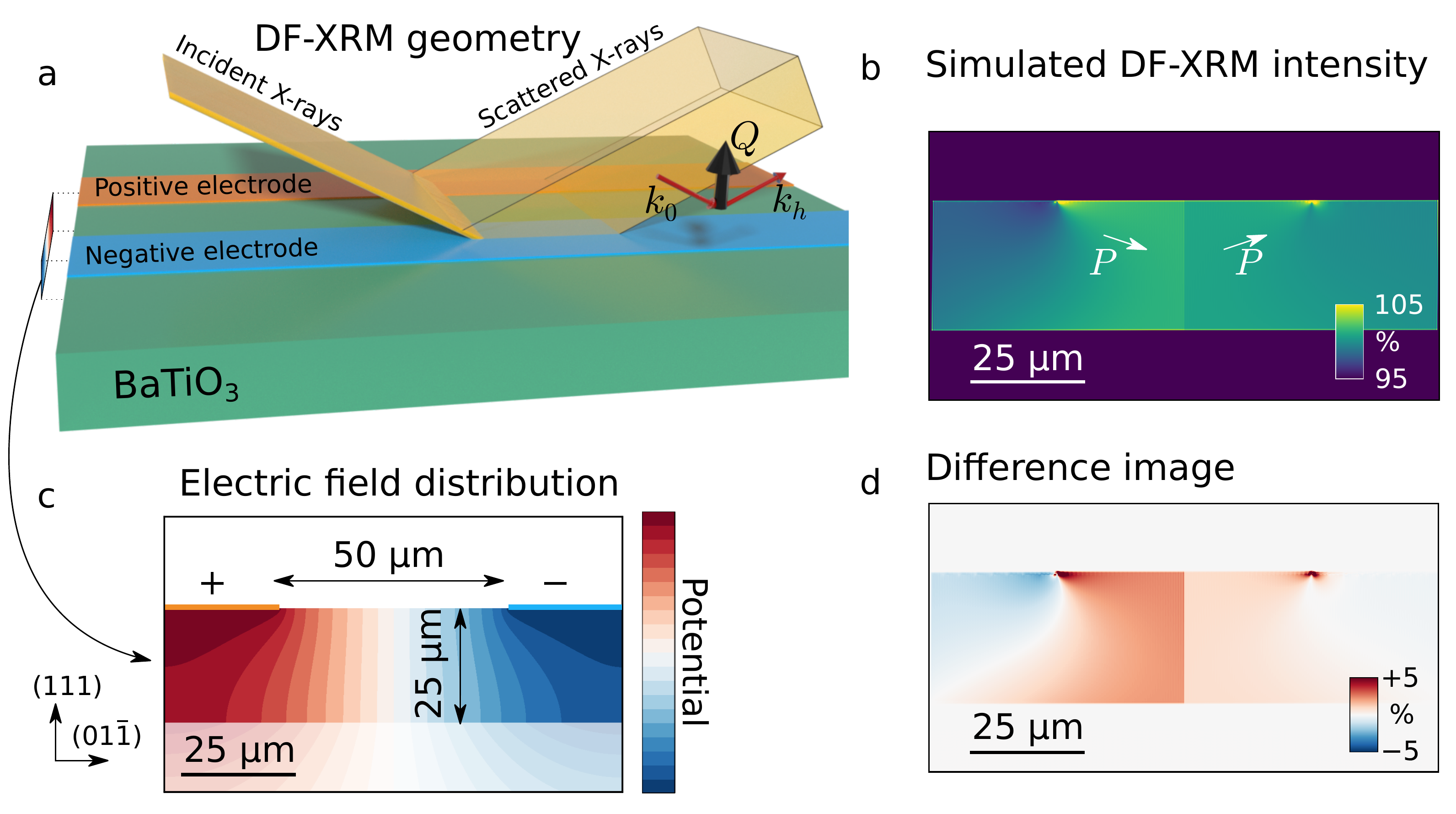}
    \caption{\textbf{a} Sketch of proposed experiment. \textbf{b} Simulated DF-XRM image  $E \sim 10$~kv/cm, $Q = (111)$. \textbf{c} Distribution of the electric potential between the positive and negative electrode. \textbf{d} Relative difference in intensity between simulated images with $E \sim 10$~kV/cm and 0~kV/cm applied between the electrodes. Subfigures b and d show the intensity distribution back-propagated to a vertical plane for comparison with subfigure c.}
    \label{fig:sim2}
\end{figure}

From Table \ref{tab:dft} one may observe that the $(011)$ and $(111)$ reflections have more than one order of magnitude higher sensitivity in intensity than the $(001)$ reflection. Imaging the electric field in BaTiO$_3$ via the intensity is then most favourable if the crystal is oriented with $\textbf{E}||\textbf{Q}=(111)$, but in order to reach 1~\% change in intensity, fields larger than the coercive field ($\sim$1~kV/cm) will be required. 
%With reasonable fields of 10~kV/cm the required noise floor is less than $1\%$ of the total intensity.
We propose a proof-of-concept experiment shown in Figure \ref{fig:sim2}a, where a slit-beam intersects a (111) oriented BaTiO$_3$ single crystal, and co-planar electrodes are utilized to generate a non-uniform electric field.
The electric field beneath the parallel electrodes is shown in Figure \ref{fig:sim2}c, while simulated DF-XRM contrast of the electric field is shown in Figure \ref{fig:sim2}b. 
Two domains are simulated ([0-10] and [001]), bisecting the sample at the center. There is no strain associated with the domain wall, and the domain configuration is stable for arbitrarily large fields. 

The simulation assumes kinematical scattering and are otherwise based on \cite{carlsen2022finite, carlsen2022simulating} with a Gauss-Schell beam\cite{starikov1982coherent} of height 0.6 µm and divergence of 0.1 mrad. %ignores dynamical diffraction to approximate the use of an X-ray diffuser\cite{falch2019experimental}.
The electric field is concentrated at the edge of the electrode and the highest electric-field induced contrast is found here. Both the in-plane and out-of-plane component of the electric field affect the scattering intensity. The effect of the in-plane electric field(orthogonal to $\textbf{Q}$) depends on the polarization direction, i.e. the ferroelectric domain. This causes the shift in contrast between the left and right side of Figure \ref{fig:sim2}b and d. Dynamical diffraction will perturb the signal if the experimental design is implemented as shown on a large, ideal sample, but the effects of dynamical diffraction may be suppressed by defects in the sample, down-scaling the geometry, or using a multi-wavelength beam.

\subsection{Experimental noise}
The results in Table \ref{tab:dft} and simulations in Figure \ref{fig:sim2} indicate a noise threshold of a few percent of the intensity.
For comparison we have analyzed a series of 100 identical DF-XRM images from a SrTiO$_3$ sample and found a standard deviation in intensity of $\sim$2-8\%, but this can be reduced to $<$1\% by simple noise-reduction methods, such as a filtering. 
In order to produce an image of the electric field, data collected in an $on$ state can be compared to data collected in an $off$ state, as illustrated in Figure \ref{fig:sim2}d. Such a comparison allows for much greater sensitivity in realistic samples where other effects may also perturb the scattered intensity. However, such comparisons are subject to drift, and a stroboscopic measurement scheme, where both $on$ and $off$ data are collected in an alternating fashion while reciprocal space is scanned, should be considered.
It follows from Equation~\ref{eq:fin_f2} that increasing the dielectric constant will increase the signal-to-noise ratio, and for BaTiO$_3$ doping and measuring closer to the phase transition at 120~$^\circ$C should be considered. 

%We have here assumed a constant dielectric constant, but as we are now in a super-coercive field regime, it is the field-dependent dielectric constant should for analysis . 
\subsection{Trends in BaTiO$_3$ and other material systems}
Equation~\ref{eq:fin_f2} suggests that any material with a dielectric constant comparable to that of BaTiO$_3$ is a candidate. 
Materials with a much lower dielectric constant (of which there are many) are effectively disqualified as dielectric breakdown will be reached before detectable ionic displacements are induced by the electric field. 
Going via the phase under the same conditions ($\textbf{E}||\textbf{Q}=(111)$, 10~kV/cm) would require the error in phase to be less than 0.5$^\circ$. This appears to be a more challenging route as the phase-recovery algorithm represents an additional source of error. The fact that two independent relationships between the phase, $\textbf{E}$, and $\textbf{u}$ must be obtained to deconvolute the strain and electric field exacerbates this issue.

Imaging in the sub-coercive field regime ($<$1~kV/cm) does not appear feasible for BaTiO$_3$. Good ferroelectric candidate materials for sub-coercive field measurements would have a high dielectric constant and a high coercive field. 
Unfortunately, the coercive field tends to decrease when the dielectric constant increases\cite{tickoo2002dielectric,shrout2007lead}.

\subsection{Implications for strain-maps of functional dielectrics}
The electric field may inadvertently influence other types of measurements in X-ray microscopy, for example strain maps. For dielectric materials with a low dielectric constant Equation~\ref{eq:fin_f2} and \ref{eq:fin_phi} imply that the perturbation of the X-ray microscopy image due to the electric field will be very low at reasonable fields. 
We suggest that the electric field is simply ignored when this is the case.

Even for materials with a higher dielectric constant, this paper describes how one may find a reflection where the contribution from the electric field is negligible. Piezoelectrics are a class of materials where this becomes particularly relevant, as some of them have high dielectric constant and strain maps are of particular interest in-situ with applied electric fields. 

We note that in ferroelectrics with multiple domains, the ferroelectric polarization cannot be ignored even if the electric field can. % as illustrated in Figure~\ref{fig:sim},
This is because the ferroelectric polarization is associated with much larger atomic displacements. 
Describing the difference in intensity and phase between different ferroelectric domains is beyond the scope of this work. However, we note that deriving this becomes much easier once it is clear that the electric field can be ignored, and the phase shift and change in intensity of a domain wall can in this case be derived from Equation~\ref{eq:scattering}.

\section{Conclusion}
This work assesses the possibility to recover an image of the electric field in X-ray microscopy. 
It has been shown that this is only possible in materials with a high dielectric constant, and then only at high electric fields and with the correct alignment. 
%Furthermore it is most promising to base the analysis on intensity rather than the phase of the scattered field, as this makes it easier to obtain the required the signal-to-noise ratio. 
Nevertheless, we propose and simulate a DF-XRM experiment where the electric field may be imaged.
Our proposed experiment is based on the well known piezo- and ferroelectric material BaTiO$_3$, which belongs to a family of ferroelectric perovskites including SrTiO$_3$, PbZr$_{x}$Ti$_{1-x}$O$_3$, and related compounds with large dielectric permittivity and strong piezoelectric and electro-optic effects, and where the behaviour of oxygen vacancies, dislocations and various dopants have been extensively studied\cite{lines2001principles,de2021transport, panda2009environmental}.
Thus, while the requirement of a high dielectric constant may limit the choice of materials, it is still remains possible to study a wide range of physical properties in relation to the shape of the electric field.

%our results show that it is possible to map the electric field in several of the materials where the relationship between strain, defects and the electric field is most critical.

%BaTiO$_3$ is the most common material for multilayer ceramic capcitors. The same material family contains PZT (PbZr$_x$Ti$_{1-x}$O$_3$) and its derivatives, used for piezoelectric applications and antiferroelectric capacitors.

%The simulations imply that it is possible to quantitatively analyze the strain fields (including those induced by the piezoelectric effect) in most functional materials by ignoring the effect of the electric field.

\begin{acknowledgments}
T. M. R. has received funding from the European Union’s Horizon 2020 research and innovation programme under the Marie Skłodowska-Curie grant agreement no. 899987. H. S. acknowledges financial support from ERC Starting Grant no. 804665. U. P. and T. O. were supported by the Villum foundation, Grant no. 00028145.
\end{acknowledgments}

\bibliography{bib}

\begin{thebibliography}{28}%
\makeatletter
\providecommand \@ifxundefined [1]{%
 \@ifx{#1\undefined}
}%
\providecommand \@ifnum [1]{%
 \ifnum #1\expandafter \@firstoftwo
 \else \expandafter \@secondoftwo
 \fi
}%
\providecommand \@ifx [1]{%
 \ifx #1\expandafter \@firstoftwo
 \else \expandafter \@secondoftwo
 \fi
}%
\providecommand \natexlab [1]{#1}%
\providecommand \enquote  [1]{``#1''}%
\providecommand \bibnamefont  [1]{#1}%
\providecommand \bibfnamefont [1]{#1}%
\providecommand \citenamefont [1]{#1}%
\providecommand \href@noop [0]{\@secondoftwo}%
\providecommand \href [0]{\begingroup \@sanitize@url \@href}%
\providecommand \@href[1]{\@@startlink{#1}\@@href}%
\providecommand \@@href[1]{\endgroup#1\@@endlink}%
\providecommand \@sanitize@url [0]{\catcode `\\12\catcode `\$12\catcode
  `\&12\catcode `\#12\catcode `\^12\catcode `\_12\catcode `\%12\relax}%
\providecommand \@@startlink[1]{}%
\providecommand \@@endlink[0]{}%
\providecommand \url  [0]{\begingroup\@sanitize@url \@url }%
\providecommand \@url [1]{\endgroup\@href {#1}{\urlprefix }}%
\providecommand \urlprefix  [0]{URL }%
\providecommand \Eprint [0]{\href }%
\providecommand \doibase [0]{https://doi.org/}%
\providecommand \selectlanguage [0]{\@gobble}%
\providecommand \bibinfo  [0]{\@secondoftwo}%
\providecommand \bibfield  [0]{\@secondoftwo}%
\providecommand \translation [1]{[#1]}%
\providecommand \BibitemOpen [0]{}%
\providecommand \bibitemStop [0]{}%
\providecommand \bibitemNoStop [0]{.\EOS\space}%
\providecommand \EOS [0]{\spacefactor3000\relax}%
\providecommand \BibitemShut  [1]{\csname bibitem#1\endcsname}%
\let\auto@bib@innerbib\@empty
%</preamble>
\bibitem [{\citenamefont {Lines}\ and\ \citenamefont
  {Glass}(2001)}]{lines2001principles}%
  \BibitemOpen
  \bibfield  {author} {\bibinfo {author} {\bibfnamefont {M.~E.}\ \bibnamefont
  {Lines}}\ and\ \bibinfo {author} {\bibfnamefont {A.~M.}\ \bibnamefont
  {Glass}},\ }\href@noop {} {\emph {\bibinfo {title} {Principles and
  applications of ferroelectrics and related materials}}}\ (\bibinfo
  {publisher} {Oxford university press},\ \bibinfo {year} {2001})\BibitemShut
  {NoStop}%
\bibitem [{\citenamefont {Pak}(1990)}]{pak1990force}%
  \BibitemOpen
  \bibfield  {author} {\bibinfo {author} {\bibfnamefont {Y.~E.}\ \bibnamefont
  {Pak}},\ }\bibfield  {title} {\bibinfo {title} {{Force on a Piezoelectric
  Screw Dislocation}},\ }\href {https://doi.org/10.1115/1.2897653} {\bibfield
  {journal} {\bibinfo  {journal} {Journal of Applied Mechanics}\ }\textbf
  {\bibinfo {volume} {57}},\ \bibinfo {pages} {863} (\bibinfo {year}
  {1990})}\BibitemShut {NoStop}%
\bibitem [{\citenamefont {Hachtel}\ \emph {et~al.}(2018)\citenamefont
  {Hachtel}, \citenamefont {Idrobo},\ and\ \citenamefont
  {Chi}}]{hachtel2018sub}%
  \BibitemOpen
  \bibfield  {author} {\bibinfo {author} {\bibfnamefont {J.~A.}\ \bibnamefont
  {Hachtel}}, \bibinfo {author} {\bibfnamefont {J.~C.}\ \bibnamefont
  {Idrobo}},\ and\ \bibinfo {author} {\bibfnamefont {M.}~\bibnamefont {Chi}},\
  }\bibfield  {title} {\bibinfo {title} {Sub-{\aa}ngstrom electric field
  measurements on a universal detector in a scanning transmission electron
  microscope},\ }\href@noop {} {\bibfield  {journal} {\bibinfo  {journal}
  {Advanced structural and chemical imaging}\ }\textbf {\bibinfo {volume}
  {4}},\ \bibinfo {pages} {1} (\bibinfo {year} {2018})}\BibitemShut {NoStop}%
\bibitem [{\citenamefont {Shibata}\ \emph {et~al.}(2015)\citenamefont
  {Shibata}, \citenamefont {Findlay}, \citenamefont {Sasaki}, \citenamefont
  {Matsumoto}, \citenamefont {Sawada}, \citenamefont {Kohno}, \citenamefont
  {Otomo}, \citenamefont {Minato},\ and\ \citenamefont
  {Ikuhara}}]{shibata2015imaging}%
  \BibitemOpen
  \bibfield  {author} {\bibinfo {author} {\bibfnamefont {N.}~\bibnamefont
  {Shibata}}, \bibinfo {author} {\bibfnamefont {S.~D.}\ \bibnamefont
  {Findlay}}, \bibinfo {author} {\bibfnamefont {H.}~\bibnamefont {Sasaki}},
  \bibinfo {author} {\bibfnamefont {T.}~\bibnamefont {Matsumoto}}, \bibinfo
  {author} {\bibfnamefont {H.}~\bibnamefont {Sawada}}, \bibinfo {author}
  {\bibfnamefont {Y.}~\bibnamefont {Kohno}}, \bibinfo {author} {\bibfnamefont
  {S.}~\bibnamefont {Otomo}}, \bibinfo {author} {\bibfnamefont
  {R.}~\bibnamefont {Minato}},\ and\ \bibinfo {author} {\bibfnamefont
  {Y.}~\bibnamefont {Ikuhara}},\ }\bibfield  {title} {\bibinfo {title} {Imaging
  of built-in electric field at a pn junction by scanning transmission electron
  microscopy},\ }\href@noop {} {\bibfield  {journal} {\bibinfo  {journal}
  {Scientific reports}\ }\textbf {\bibinfo {volume} {5}},\ \bibinfo {pages}
  {10040} (\bibinfo {year} {2015})}\BibitemShut {NoStop}%
\bibitem [{\citenamefont {Melitz}\ \emph {et~al.}(2011)\citenamefont {Melitz},
  \citenamefont {Shen}, \citenamefont {Kummel},\ and\ \citenamefont
  {Lee}}]{melitz2011kelvin}%
  \BibitemOpen
  \bibfield  {author} {\bibinfo {author} {\bibfnamefont {W.}~\bibnamefont
  {Melitz}}, \bibinfo {author} {\bibfnamefont {J.}~\bibnamefont {Shen}},
  \bibinfo {author} {\bibfnamefont {A.~C.}\ \bibnamefont {Kummel}},\ and\
  \bibinfo {author} {\bibfnamefont {S.}~\bibnamefont {Lee}},\ }\bibfield
  {title} {\bibinfo {title} {Kelvin probe force microscopy and its
  application},\ }\href@noop {} {\bibfield  {journal} {\bibinfo  {journal}
  {Surface science reports}\ }\textbf {\bibinfo {volume} {66}},\ \bibinfo
  {pages} {1} (\bibinfo {year} {2011})}\BibitemShut {NoStop}%
\bibitem [{\citenamefont {Simons}\ \emph {et~al.}(2015)\citenamefont {Simons},
  \citenamefont {King}, \citenamefont {Ludwig}, \citenamefont {Detlefs},
  \citenamefont {Pantleon}, \citenamefont {Schmidt}, \citenamefont {St{\"o}hr},
  \citenamefont {Snigireva}, \citenamefont {Snigirev},\ and\ \citenamefont
  {Poulsen}}]{simons2015dark}%
  \BibitemOpen
  \bibfield  {author} {\bibinfo {author} {\bibfnamefont {H.}~\bibnamefont
  {Simons}}, \bibinfo {author} {\bibfnamefont {A.}~\bibnamefont {King}},
  \bibinfo {author} {\bibfnamefont {W.}~\bibnamefont {Ludwig}}, \bibinfo
  {author} {\bibfnamefont {C.}~\bibnamefont {Detlefs}}, \bibinfo {author}
  {\bibfnamefont {W.}~\bibnamefont {Pantleon}}, \bibinfo {author}
  {\bibfnamefont {S.}~\bibnamefont {Schmidt}}, \bibinfo {author} {\bibfnamefont
  {F.}~\bibnamefont {St{\"o}hr}}, \bibinfo {author} {\bibfnamefont
  {I.}~\bibnamefont {Snigireva}}, \bibinfo {author} {\bibfnamefont
  {A.}~\bibnamefont {Snigirev}},\ and\ \bibinfo {author} {\bibfnamefont
  {H.~F.}\ \bibnamefont {Poulsen}},\ }\bibfield  {title} {\bibinfo {title}
  {Dark-field x-ray microscopy for multiscale structural characterization},\
  }\href@noop {} {\bibfield  {journal} {\bibinfo  {journal} {Nature
  communications}\ }\textbf {\bibinfo {volume} {6}},\ \bibinfo {pages} {1}
  (\bibinfo {year} {2015})}\BibitemShut {NoStop}%
\bibitem [{\citenamefont {Shi}\ \emph {et~al.}(2022)\citenamefont {Shi},
  \citenamefont {Shi},\ and\ \citenamefont {Fohtung}}]{shi2022applicability}%
  \BibitemOpen
  \bibfield  {author} {\bibinfo {author} {\bibfnamefont {X.}~\bibnamefont
  {Shi}}, \bibinfo {author} {\bibfnamefont {J.}~\bibnamefont {Shi}},\ and\
  \bibinfo {author} {\bibfnamefont {E.}~\bibnamefont {Fohtung}},\ }\bibfield
  {title} {\bibinfo {title} {Applicability of coherent x-ray diffractive
  imaging to ferroelectric, ferromagnetic, and phase change materials},\
  }\href@noop {} {\bibfield  {journal} {\bibinfo  {journal} {Journal of Applied
  Physics}\ }\textbf {\bibinfo {volume} {131}},\ \bibinfo {pages} {040901}
  (\bibinfo {year} {2022})}\BibitemShut {NoStop}%
\bibitem [{\citenamefont {Chahine}\ \emph {et~al.}(2014)\citenamefont
  {Chahine}, \citenamefont {Richard}, \citenamefont {Homs-Regojo},
  \citenamefont {Tran-Caliste}, \citenamefont {Carbone}, \citenamefont
  {Jacques}, \citenamefont {Grifone}, \citenamefont {Boesecke}, \citenamefont
  {Katzer}, \citenamefont {Costina} \emph {et~al.}}]{chahine2014imaging}%
  \BibitemOpen
  \bibfield  {author} {\bibinfo {author} {\bibfnamefont {G.~A.}\ \bibnamefont
  {Chahine}}, \bibinfo {author} {\bibfnamefont {M.-I.}\ \bibnamefont
  {Richard}}, \bibinfo {author} {\bibfnamefont {R.~A.}\ \bibnamefont
  {Homs-Regojo}}, \bibinfo {author} {\bibfnamefont {T.~N.}\ \bibnamefont
  {Tran-Caliste}}, \bibinfo {author} {\bibfnamefont {D.}~\bibnamefont
  {Carbone}}, \bibinfo {author} {\bibfnamefont {V.~L.~R.}\ \bibnamefont
  {Jacques}}, \bibinfo {author} {\bibfnamefont {R.}~\bibnamefont {Grifone}},
  \bibinfo {author} {\bibfnamefont {P.}~\bibnamefont {Boesecke}}, \bibinfo
  {author} {\bibfnamefont {J.}~\bibnamefont {Katzer}}, \bibinfo {author}
  {\bibfnamefont {I.}~\bibnamefont {Costina}}, \emph {et~al.},\ }\bibfield
  {title} {\bibinfo {title} {Imaging of strain and lattice orientation by quick
  scanning x-ray microscopy combined with three-dimensional reciprocal space
  mapping},\ }\href@noop {} {\bibfield  {journal} {\bibinfo  {journal} {Journal
  of Applied Crystallography}\ }\textbf {\bibinfo {volume} {47}},\ \bibinfo
  {pages} {762} (\bibinfo {year} {2014})}\BibitemShut {NoStop}%
\bibitem [{\citenamefont {Pfeiffer}(2018)}]{pfeiffer2018x}%
  \BibitemOpen
  \bibfield  {author} {\bibinfo {author} {\bibfnamefont {F.}~\bibnamefont
  {Pfeiffer}},\ }\bibfield  {title} {\bibinfo {title} {X-ray ptychography},\
  }\href@noop {} {\bibfield  {journal} {\bibinfo  {journal} {Nature Photonics}\
  }\textbf {\bibinfo {volume} {12}},\ \bibinfo {pages} {9} (\bibinfo {year}
  {2018})}\BibitemShut {NoStop}%
\bibitem [{\citenamefont {Simons}\ \emph {et~al.}(2018)\citenamefont {Simons},
  \citenamefont {Haugen}, \citenamefont {Jakobsen}, \citenamefont {Schmidt},
  \citenamefont {St{\"o}hr}, \citenamefont {Majkut}, \citenamefont {Detlefs},
  \citenamefont {Daniels}, \citenamefont {Damjanovic},\ and\ \citenamefont
  {Poulsen}}]{simons2018long}%
  \BibitemOpen
  \bibfield  {author} {\bibinfo {author} {\bibfnamefont {H.}~\bibnamefont
  {Simons}}, \bibinfo {author} {\bibfnamefont {A.~B.}\ \bibnamefont {Haugen}},
  \bibinfo {author} {\bibfnamefont {A.~C.}\ \bibnamefont {Jakobsen}}, \bibinfo
  {author} {\bibfnamefont {S.}~\bibnamefont {Schmidt}}, \bibinfo {author}
  {\bibfnamefont {F.}~\bibnamefont {St{\"o}hr}}, \bibinfo {author}
  {\bibfnamefont {M.}~\bibnamefont {Majkut}}, \bibinfo {author} {\bibfnamefont
  {C.}~\bibnamefont {Detlefs}}, \bibinfo {author} {\bibfnamefont {J.~E.}\
  \bibnamefont {Daniels}}, \bibinfo {author} {\bibfnamefont {D.}~\bibnamefont
  {Damjanovic}},\ and\ \bibinfo {author} {\bibfnamefont {H.~F.}\ \bibnamefont
  {Poulsen}},\ }\bibfield  {title} {\bibinfo {title} {Long-range symmetry
  breaking in embedded ferroelectrics},\ }\href@noop {} {\bibfield  {journal}
  {\bibinfo  {journal} {Nature materials}\ }\textbf {\bibinfo {volume} {17}},\
  \bibinfo {pages} {814} (\bibinfo {year} {2018})}\BibitemShut {NoStop}%
\bibitem [{\citenamefont {Gorfman}\ \emph {et~al.}(2016)\citenamefont
  {Gorfman}, \citenamefont {Simons}, \citenamefont {Iamsasri}, \citenamefont
  {Prasertpalichat}, \citenamefont {Cann}, \citenamefont {Choe}, \citenamefont
  {Pietsch}, \citenamefont {Watier},\ and\ \citenamefont
  {Jones}}]{gorfman2016simultaneous}%
  \BibitemOpen
  \bibfield  {author} {\bibinfo {author} {\bibfnamefont {S.}~\bibnamefont
  {Gorfman}}, \bibinfo {author} {\bibfnamefont {H.}~\bibnamefont {Simons}},
  \bibinfo {author} {\bibfnamefont {T.}~\bibnamefont {Iamsasri}}, \bibinfo
  {author} {\bibfnamefont {S.}~\bibnamefont {Prasertpalichat}}, \bibinfo
  {author} {\bibfnamefont {D.}~\bibnamefont {Cann}}, \bibinfo {author}
  {\bibfnamefont {H.}~\bibnamefont {Choe}}, \bibinfo {author} {\bibfnamefont
  {U.}~\bibnamefont {Pietsch}}, \bibinfo {author} {\bibfnamefont
  {Y.}~\bibnamefont {Watier}},\ and\ \bibinfo {author} {\bibfnamefont
  {J.}~\bibnamefont {Jones}},\ }\bibfield  {title} {\bibinfo {title}
  {Simultaneous resonant x-ray diffraction measurement of polarization
  inversion and lattice strain in polycrystalline ferroelectrics},\ }\href@noop
  {} {\bibfield  {journal} {\bibinfo  {journal} {Scientific reports}\ }\textbf
  {\bibinfo {volume} {6}},\ \bibinfo {pages} {1} (\bibinfo {year}
  {2016})}\BibitemShut {NoStop}%
\bibitem [{\citenamefont {Grigoriev}\ \emph {et~al.}(2006)\citenamefont
  {Grigoriev}, \citenamefont {Do}, \citenamefont {Kim}, \citenamefont {Eom},
  \citenamefont {Evans}, \citenamefont {Adams},\ and\ \citenamefont
  {Dufresne}}]{grigoriev2006nanosecond}%
  \BibitemOpen
  \bibfield  {author} {\bibinfo {author} {\bibfnamefont {A.}~\bibnamefont
  {Grigoriev}}, \bibinfo {author} {\bibfnamefont {D.-H.}\ \bibnamefont {Do}},
  \bibinfo {author} {\bibfnamefont {D.~M.}\ \bibnamefont {Kim}}, \bibinfo
  {author} {\bibfnamefont {C.-B.}\ \bibnamefont {Eom}}, \bibinfo {author}
  {\bibfnamefont {P.~G.}\ \bibnamefont {Evans}}, \bibinfo {author}
  {\bibfnamefont {B.~W.}\ \bibnamefont {Adams}},\ and\ \bibinfo {author}
  {\bibfnamefont {E.~M.}\ \bibnamefont {Dufresne}},\ }\bibfield  {title}
  {\bibinfo {title} {Nanosecond structural visualization of the reproducibility
  of polarization switching in ferroelectrics},\ }\href@noop {} {\bibfield
  {journal} {\bibinfo  {journal} {Integrated Ferroelectrics}\ }\textbf
  {\bibinfo {volume} {85}},\ \bibinfo {pages} {165} (\bibinfo {year}
  {2006})}\BibitemShut {NoStop}%
\bibitem [{\citenamefont {Hruszkewycz}\ \emph {et~al.}(2013)\citenamefont
  {Hruszkewycz}, \citenamefont {Highland}, \citenamefont {Holt}, \citenamefont
  {Kim}, \citenamefont {Folkman}, \citenamefont {Thompson}, \citenamefont
  {Tripathi}, \citenamefont {Stephenson}, \citenamefont {Hong},\ and\
  \citenamefont {Fuoss}}]{hruszkewycz2013imaging}%
  \BibitemOpen
  \bibfield  {author} {\bibinfo {author} {\bibfnamefont {S.}~\bibnamefont
  {Hruszkewycz}}, \bibinfo {author} {\bibfnamefont {M.}~\bibnamefont
  {Highland}}, \bibinfo {author} {\bibfnamefont {M.}~\bibnamefont {Holt}},
  \bibinfo {author} {\bibfnamefont {D.}~\bibnamefont {Kim}}, \bibinfo {author}
  {\bibfnamefont {C.}~\bibnamefont {Folkman}}, \bibinfo {author} {\bibfnamefont
  {C.}~\bibnamefont {Thompson}}, \bibinfo {author} {\bibfnamefont
  {A.}~\bibnamefont {Tripathi}}, \bibinfo {author} {\bibfnamefont
  {G.}~\bibnamefont {Stephenson}}, \bibinfo {author} {\bibfnamefont
  {S.}~\bibnamefont {Hong}},\ and\ \bibinfo {author} {\bibfnamefont
  {P.}~\bibnamefont {Fuoss}},\ }\bibfield  {title} {\bibinfo {title} {Imaging
  local polarization in ferroelectric thin films by coherent x-ray bragg
  projection ptychography},\ }\href@noop {} {\bibfield  {journal} {\bibinfo
  {journal} {Physical Review Letters}\ }\textbf {\bibinfo {volume} {110}},\
  \bibinfo {pages} {177601} (\bibinfo {year} {2013})}\BibitemShut {NoStop}%
\bibitem [{\citenamefont {Shabalin}\ and\ \citenamefont
  {Shpyrko}(2021)}]{shabalin2021multiwavelength}%
  \BibitemOpen
  \bibfield  {author} {\bibinfo {author} {\bibfnamefont {A.~G.}\ \bibnamefont
  {Shabalin}}\ and\ \bibinfo {author} {\bibfnamefont {O.~G.}\ \bibnamefont
  {Shpyrko}},\ }\bibfield  {title} {\bibinfo {title} {Multiwavelength anomalous
  x-ray diffraction for combined imaging of atomic displacement and strain},\
  }\href@noop {} {\bibfield  {journal} {\bibinfo  {journal} {Acta
  Crystallographica Section A: Foundations and Advances}\ }\textbf {\bibinfo
  {volume} {77}} (\bibinfo {year} {2021})}\BibitemShut {NoStop}%
\bibitem [{\citenamefont {Carlsen}\ \emph
  {et~al.}(2022{\natexlab{a}})\citenamefont {Carlsen}, \citenamefont
  {R{\ae}der}, \citenamefont {Yildirim}, \citenamefont {Rodriguez-Lamas},
  \citenamefont {Detlefs},\ and\ \citenamefont {Simons}}]{mads_phase_recovery}%
  \BibitemOpen
  \bibfield  {author} {\bibinfo {author} {\bibfnamefont {M.}~\bibnamefont
  {Carlsen}}, \bibinfo {author} {\bibfnamefont {T.~M.}\ \bibnamefont
  {R{\ae}der}}, \bibinfo {author} {\bibfnamefont {C.}~\bibnamefont {Yildirim}},
  \bibinfo {author} {\bibfnamefont {R.}~\bibnamefont {Rodriguez-Lamas}},
  \bibinfo {author} {\bibfnamefont {C.}~\bibnamefont {Detlefs}},\ and\ \bibinfo
  {author} {\bibfnamefont {H.}~\bibnamefont {Simons}},\ }\bibfield  {title}
  {\bibinfo {title} {Fourier ptychographic dark field x-ray microscopy},\
  }\href@noop {} {\bibfield  {journal} {\bibinfo  {journal} {Optics Express}\
  }\textbf {\bibinfo {volume} {30}},\ \bibinfo {pages} {2949} (\bibinfo {year}
  {2022}{\natexlab{a}})}\BibitemShut {NoStop}%
\bibitem [{\citenamefont {Roux}\ and\ \citenamefont {del
  Rio}()}]{dabaxdynamic}%
  \BibitemOpen
  \bibfield  {author} {\bibinfo {author} {\bibfnamefont {B.}~\bibnamefont
  {Roux}}\ and\ \bibinfo {author} {\bibfnamefont {M.}~\bibnamefont {del Rio}},\
  }\href@noop {} {\emph {\bibinfo {title} {DABAX, DAtaBAse for X-ray
  applications}}}\ (\bibinfo  {publisher} {European Synchrotron Radiation
  Facility})\BibitemShut {NoStop}%
\bibitem [{\citenamefont {Larsen}\ \emph {et~al.}(2017)\citenamefont {Larsen},
  \citenamefont {Mortensen}, \citenamefont {Blomqvist}, \citenamefont
  {Castelli}, \citenamefont {Christensen}, \citenamefont {Du{\l}ak},
  \citenamefont {Friis}, \citenamefont {Groves}, \citenamefont {Hammer},
  \citenamefont {Hargus}, \citenamefont {Hermes}, \citenamefont {Jennings},
  \citenamefont {Jensen}, \citenamefont {Kermode}, \citenamefont {Kitchin},
  \citenamefont {Kolsbjerg}, \citenamefont {Kubal}, \citenamefont {Kaasbjerg},
  \citenamefont {Lysgaard}, \citenamefont {Maronsson}, \citenamefont {Maxson},
  \citenamefont {Olsen}, \citenamefont {Pastewka}, \citenamefont {Peterson},
  \citenamefont {Rostgaard}, \citenamefont {Schi{\o}tz}, \citenamefont
  {Schütt}, \citenamefont {Strange}, \citenamefont {Thygesen}, \citenamefont
  {Vegge}, \citenamefont {Vilhelmsen}, \citenamefont {Walter}, \citenamefont
  {Zeng},\ and\ \citenamefont {Jacobsen}}]{Hjorth_Larsen_2017}%
  \BibitemOpen
  \bibfield  {author} {\bibinfo {author} {\bibfnamefont {A.~H.}\ \bibnamefont
  {Larsen}}, \bibinfo {author} {\bibfnamefont {J.~J.}\ \bibnamefont
  {Mortensen}}, \bibinfo {author} {\bibfnamefont {J.}~\bibnamefont
  {Blomqvist}}, \bibinfo {author} {\bibfnamefont {I.~E.}\ \bibnamefont
  {Castelli}}, \bibinfo {author} {\bibfnamefont {R.}~\bibnamefont
  {Christensen}}, \bibinfo {author} {\bibfnamefont {M.}~\bibnamefont
  {Du{\l}ak}}, \bibinfo {author} {\bibfnamefont {J.}~\bibnamefont {Friis}},
  \bibinfo {author} {\bibfnamefont {M.~N.}\ \bibnamefont {Groves}}, \bibinfo
  {author} {\bibfnamefont {B.}~\bibnamefont {Hammer}}, \bibinfo {author}
  {\bibfnamefont {C.}~\bibnamefont {Hargus}}, \bibinfo {author} {\bibfnamefont
  {E.~D.}\ \bibnamefont {Hermes}}, \bibinfo {author} {\bibfnamefont {P.~C.}\
  \bibnamefont {Jennings}}, \bibinfo {author} {\bibfnamefont {P.~B.}\
  \bibnamefont {Jensen}}, \bibinfo {author} {\bibfnamefont {J.}~\bibnamefont
  {Kermode}}, \bibinfo {author} {\bibfnamefont {J.~R.}\ \bibnamefont
  {Kitchin}}, \bibinfo {author} {\bibfnamefont {E.~L.}\ \bibnamefont
  {Kolsbjerg}}, \bibinfo {author} {\bibfnamefont {J.}~\bibnamefont {Kubal}},
  \bibinfo {author} {\bibfnamefont {K.}~\bibnamefont {Kaasbjerg}}, \bibinfo
  {author} {\bibfnamefont {S.}~\bibnamefont {Lysgaard}}, \bibinfo {author}
  {\bibfnamefont {J.~B.}\ \bibnamefont {Maronsson}}, \bibinfo {author}
  {\bibfnamefont {T.}~\bibnamefont {Maxson}}, \bibinfo {author} {\bibfnamefont
  {T.}~\bibnamefont {Olsen}}, \bibinfo {author} {\bibfnamefont
  {L.}~\bibnamefont {Pastewka}}, \bibinfo {author} {\bibfnamefont
  {A.}~\bibnamefont {Peterson}}, \bibinfo {author} {\bibfnamefont
  {C.}~\bibnamefont {Rostgaard}}, \bibinfo {author} {\bibfnamefont
  {J.}~\bibnamefont {Schi{\o}tz}}, \bibinfo {author} {\bibfnamefont
  {O.}~\bibnamefont {Schütt}}, \bibinfo {author} {\bibfnamefont
  {M.}~\bibnamefont {Strange}}, \bibinfo {author} {\bibfnamefont {K.~S.}\
  \bibnamefont {Thygesen}}, \bibinfo {author} {\bibfnamefont {T.}~\bibnamefont
  {Vegge}}, \bibinfo {author} {\bibfnamefont {L.}~\bibnamefont {Vilhelmsen}},
  \bibinfo {author} {\bibfnamefont {M.}~\bibnamefont {Walter}}, \bibinfo
  {author} {\bibfnamefont {Z.}~\bibnamefont {Zeng}},\ and\ \bibinfo {author}
  {\bibfnamefont {K.~W.}\ \bibnamefont {Jacobsen}},\ }\bibfield  {title}
  {\bibinfo {title} {The atomic simulation environment{\textemdash}a python
  library for working with atoms},\ }\href
  {https://doi.org/10.1088/1361-648x/aa680e} {\bibfield  {journal} {\bibinfo
  {journal} {Journal of Physics: Condensed Matter}\ }\textbf {\bibinfo {volume}
  {29}},\ \bibinfo {pages} {273002} (\bibinfo {year} {2017})}\BibitemShut
  {NoStop}%
\bibitem [{\citenamefont {Enkovaara}\ \emph {et~al.}(2010)\citenamefont
  {Enkovaara}, \citenamefont {Rostgaard}, \citenamefont {Mortensen},
  \citenamefont {Chen}, \citenamefont {Du{\l}ak}, \citenamefont {Ferrighi},
  \citenamefont {Gavnholt}, \citenamefont {Glinsvad}, \citenamefont {Haikola},
  \citenamefont {Hansen}, \citenamefont {Kristoffersen}, \citenamefont
  {Kuisma}, \citenamefont {Larsen}, \citenamefont {Lehtovaara}, \citenamefont
  {Ljungberg}, \citenamefont {Lopez-Acevedo}, \citenamefont {Moses},
  \citenamefont {Ojanen}, \citenamefont {Olsen}, \citenamefont {Petzold},
  \citenamefont {Romero}, \citenamefont {Stausholm-M{\o}ller}, \citenamefont
  {Strange}, \citenamefont {Tritsaris}, \citenamefont {Vanin}, \citenamefont
  {Walter}, \citenamefont {Hammer}, \citenamefont {Häkkinen}, \citenamefont
  {Madsen}, \citenamefont {Nieminen}, \citenamefont {N{\o}rskov}, \citenamefont
  {Puska}, \citenamefont {Rantala}, \citenamefont {Schi{\o}tz}, \citenamefont
  {Thygesen},\ and\ \citenamefont {Jacobsen}}]{Enkovaara_2010}%
  \BibitemOpen
  \bibfield  {author} {\bibinfo {author} {\bibfnamefont {J.}~\bibnamefont
  {Enkovaara}}, \bibinfo {author} {\bibfnamefont {C.}~\bibnamefont
  {Rostgaard}}, \bibinfo {author} {\bibfnamefont {J.~J.}\ \bibnamefont
  {Mortensen}}, \bibinfo {author} {\bibfnamefont {J.}~\bibnamefont {Chen}},
  \bibinfo {author} {\bibfnamefont {M.}~\bibnamefont {Du{\l}ak}}, \bibinfo
  {author} {\bibfnamefont {L.}~\bibnamefont {Ferrighi}}, \bibinfo {author}
  {\bibfnamefont {J.}~\bibnamefont {Gavnholt}}, \bibinfo {author}
  {\bibfnamefont {C.}~\bibnamefont {Glinsvad}}, \bibinfo {author}
  {\bibfnamefont {V.}~\bibnamefont {Haikola}}, \bibinfo {author} {\bibfnamefont
  {H.~A.}\ \bibnamefont {Hansen}}, \bibinfo {author} {\bibfnamefont {H.~H.}\
  \bibnamefont {Kristoffersen}}, \bibinfo {author} {\bibfnamefont
  {M.}~\bibnamefont {Kuisma}}, \bibinfo {author} {\bibfnamefont {A.~H.}\
  \bibnamefont {Larsen}}, \bibinfo {author} {\bibfnamefont {L.}~\bibnamefont
  {Lehtovaara}}, \bibinfo {author} {\bibfnamefont {M.}~\bibnamefont
  {Ljungberg}}, \bibinfo {author} {\bibfnamefont {O.}~\bibnamefont
  {Lopez-Acevedo}}, \bibinfo {author} {\bibfnamefont {P.~G.}\ \bibnamefont
  {Moses}}, \bibinfo {author} {\bibfnamefont {J.}~\bibnamefont {Ojanen}},
  \bibinfo {author} {\bibfnamefont {T.}~\bibnamefont {Olsen}}, \bibinfo
  {author} {\bibfnamefont {V.}~\bibnamefont {Petzold}}, \bibinfo {author}
  {\bibfnamefont {N.~A.}\ \bibnamefont {Romero}}, \bibinfo {author}
  {\bibfnamefont {J.}~\bibnamefont {Stausholm-M{\o}ller}}, \bibinfo {author}
  {\bibfnamefont {M.}~\bibnamefont {Strange}}, \bibinfo {author} {\bibfnamefont
  {G.~A.}\ \bibnamefont {Tritsaris}}, \bibinfo {author} {\bibfnamefont
  {M.}~\bibnamefont {Vanin}}, \bibinfo {author} {\bibfnamefont
  {M.}~\bibnamefont {Walter}}, \bibinfo {author} {\bibfnamefont
  {B.}~\bibnamefont {Hammer}}, \bibinfo {author} {\bibfnamefont
  {H.}~\bibnamefont {Häkkinen}}, \bibinfo {author} {\bibfnamefont {G.~K.~H.}\
  \bibnamefont {Madsen}}, \bibinfo {author} {\bibfnamefont {R.~M.}\
  \bibnamefont {Nieminen}}, \bibinfo {author} {\bibfnamefont {J.~K.}\
  \bibnamefont {N{\o}rskov}}, \bibinfo {author} {\bibfnamefont
  {M.}~\bibnamefont {Puska}}, \bibinfo {author} {\bibfnamefont {T.~T.}\
  \bibnamefont {Rantala}}, \bibinfo {author} {\bibfnamefont {J.}~\bibnamefont
  {Schi{\o}tz}}, \bibinfo {author} {\bibfnamefont {K.~S.}\ \bibnamefont
  {Thygesen}},\ and\ \bibinfo {author} {\bibfnamefont {K.~W.}\ \bibnamefont
  {Jacobsen}},\ }\bibfield  {title} {\bibinfo {title} {Electronic structure
  calculations with {GPAW}: a real-space implementation of the projector
  augmented-wave method},\ }\href
  {https://doi.org/10.1088/0953-8984/22/25/253202} {\bibfield  {journal}
  {\bibinfo  {journal} {Journal of Physics: Condensed Matter}\ }\textbf
  {\bibinfo {volume} {22}},\ \bibinfo {pages} {253202} (\bibinfo {year}
  {2010})}\BibitemShut {NoStop}%
\bibitem [{\citenamefont {Togo}\ and\ \citenamefont {Tanaka}(2015)}]{Togo2015}%
  \BibitemOpen
  \bibfield  {author} {\bibinfo {author} {\bibfnamefont {A.}~\bibnamefont
  {Togo}}\ and\ \bibinfo {author} {\bibfnamefont {I.}~\bibnamefont {Tanaka}},\
  }\bibfield  {title} {\bibinfo {title} {First principles phonon calculations
  in materials science},\ }\href
  {https://doi.org/https://doi.org/10.1016/j.scriptamat.2015.07.021} {\bibfield
   {journal} {\bibinfo  {journal} {Scripta Materialia}\ }\textbf {\bibinfo
  {volume} {108}},\ \bibinfo {pages} {1} (\bibinfo {year} {2015})}\BibitemShut
  {NoStop}%
\bibitem [{\citenamefont {Peng}\ \emph {et~al.}(2020)\citenamefont {Peng},
  \citenamefont {Hu}, \citenamefont {Murakami}, \citenamefont {Zhang},\ and\
  \citenamefont {Monserrat}}]{Peng2020}%
  \BibitemOpen
  \bibfield  {author} {\bibinfo {author} {\bibfnamefont {B.}~\bibnamefont
  {Peng}}, \bibinfo {author} {\bibfnamefont {Y.}~\bibnamefont {Hu}}, \bibinfo
  {author} {\bibfnamefont {S.}~\bibnamefont {Murakami}}, \bibinfo {author}
  {\bibfnamefont {T.}~\bibnamefont {Zhang}},\ and\ \bibinfo {author}
  {\bibfnamefont {B.}~\bibnamefont {Monserrat}},\ }\bibfield  {title} {\bibinfo
  {title} {Topological phonons in oxide perovskites controlled by light},\
  }\href {https://doi.org/10.1126/sciadv.abd1618} {\bibfield  {journal}
  {\bibinfo  {journal} {Science Advances}\ }\textbf {\bibinfo {volume} {6}},\
  \bibinfo {pages} {eabd1618} (\bibinfo {year} {2020})}\BibitemShut {NoStop}%
\bibitem [{\citenamefont {Merz}(1949)}]{merz1949electric}%
  \BibitemOpen
  \bibfield  {author} {\bibinfo {author} {\bibfnamefont {W.~J.}\ \bibnamefont
  {Merz}},\ }\bibfield  {title} {\bibinfo {title} {The electric and optical
  behavior of batio$_3$ single-domain crystals},\ }\href@noop {} {\bibfield
  {journal} {\bibinfo  {journal} {Physical Review}\ }\textbf {\bibinfo {volume}
  {76}},\ \bibinfo {pages} {1221} (\bibinfo {year} {1949})}\BibitemShut
  {NoStop}%
\bibitem [{\citenamefont {Carlsen}\ and\ \citenamefont
  {Simons}(2022)}]{carlsen2022finite}%
  \BibitemOpen
  \bibfield  {author} {\bibinfo {author} {\bibfnamefont {M.}~\bibnamefont
  {Carlsen}}\ and\ \bibinfo {author} {\bibfnamefont {H.}~\bibnamefont
  {Simons}},\ }\bibfield  {title} {\bibinfo {title} {A finite difference scheme
  for integrating the takagi--taupin equations on an arbitrary orthogonal
  grid},\ }\href@noop {} {\bibfield  {journal} {\bibinfo  {journal} {Acta
  Crystallographica Section A: Foundations and Advances}\ }\textbf {\bibinfo
  {volume} {78}} (\bibinfo {year} {2022})}\BibitemShut {NoStop}%
\bibitem [{\citenamefont {Carlsen}\ \emph
  {et~al.}(2022{\natexlab{b}})\citenamefont {Carlsen}, \citenamefont {Detlefs},
  \citenamefont {Yildirim}, \citenamefont {R{\ae}der},\ and\ \citenamefont
  {Simons}}]{carlsen2022simulating}%
  \BibitemOpen
  \bibfield  {author} {\bibinfo {author} {\bibfnamefont {M.}~\bibnamefont
  {Carlsen}}, \bibinfo {author} {\bibfnamefont {C.}~\bibnamefont {Detlefs}},
  \bibinfo {author} {\bibfnamefont {C.}~\bibnamefont {Yildirim}}, \bibinfo
  {author} {\bibfnamefont {T.}~\bibnamefont {R{\ae}der}},\ and\ \bibinfo
  {author} {\bibfnamefont {H.}~\bibnamefont {Simons}},\ }\bibfield  {title}
  {\bibinfo {title} {Simulating dark-field x-ray microscopy images with
  wavefront propagation techniques},\ }\href@noop {} {\bibfield  {journal}
  {\bibinfo  {journal} {Acta Crystallographica Section A: Foundations and
  Advances}\ }\textbf {\bibinfo {volume} {78}} (\bibinfo {year}
  {2022}{\natexlab{b}})}\BibitemShut {NoStop}%
\bibitem [{\citenamefont {Starikov}\ and\ \citenamefont
  {Wolf}(1982)}]{starikov1982coherent}%
  \BibitemOpen
  \bibfield  {author} {\bibinfo {author} {\bibfnamefont {A.}~\bibnamefont
  {Starikov}}\ and\ \bibinfo {author} {\bibfnamefont {E.}~\bibnamefont
  {Wolf}},\ }\bibfield  {title} {\bibinfo {title} {Coherent-mode representation
  of gaussian schell-model sources and of their radiation fields},\ }\href@noop
  {} {\bibfield  {journal} {\bibinfo  {journal} {JOSA}\ }\textbf {\bibinfo
  {volume} {72}},\ \bibinfo {pages} {923} (\bibinfo {year} {1982})}\BibitemShut
  {NoStop}%
\bibitem [{\citenamefont {Tickoo}\ \emph {et~al.}(2002)\citenamefont {Tickoo},
  \citenamefont {Tandon}, \citenamefont {Mehra},\ and\ \citenamefont
  {Kotru}}]{tickoo2002dielectric}%
  \BibitemOpen
  \bibfield  {author} {\bibinfo {author} {\bibfnamefont {R.}~\bibnamefont
  {Tickoo}}, \bibinfo {author} {\bibfnamefont {R.}~\bibnamefont {Tandon}},
  \bibinfo {author} {\bibfnamefont {N.}~\bibnamefont {Mehra}},\ and\ \bibinfo
  {author} {\bibfnamefont {P.}~\bibnamefont {Kotru}},\ }\bibfield  {title}
  {\bibinfo {title} {Dielectric and ferroelectric properties of lanthanum
  modified lead titanate ceramics},\ }\href@noop {} {\bibfield  {journal}
  {\bibinfo  {journal} {Materials Science and Engineering: B}\ }\textbf
  {\bibinfo {volume} {94}},\ \bibinfo {pages} {1} (\bibinfo {year}
  {2002})}\BibitemShut {NoStop}%
\bibitem [{\citenamefont {Shrout}\ and\ \citenamefont
  {Zhang}(2007)}]{shrout2007lead}%
  \BibitemOpen
  \bibfield  {author} {\bibinfo {author} {\bibfnamefont {T.~R.}\ \bibnamefont
  {Shrout}}\ and\ \bibinfo {author} {\bibfnamefont {S.~J.}\ \bibnamefont
  {Zhang}},\ }\bibfield  {title} {\bibinfo {title} {Lead-free piezoelectric
  ceramics: Alternatives for pzt?},\ }\href@noop {} {\bibfield  {journal}
  {\bibinfo  {journal} {Journal of Electroceramics}\ }\textbf {\bibinfo
  {volume} {19}},\ \bibinfo {pages} {113} (\bibinfo {year} {2007})}\BibitemShut
  {NoStop}%
\bibitem [{\citenamefont {De~Souza}(2021)}]{de2021transport}%
  \BibitemOpen
  \bibfield  {author} {\bibinfo {author} {\bibfnamefont {R.~A.}\ \bibnamefont
  {De~Souza}},\ }\bibfield  {title} {\bibinfo {title} {Transport properties of
  dislocations in srtio$_3$ and other perovskites},\ }\href@noop {} {\bibfield
  {journal} {\bibinfo  {journal} {Current Opinion in Solid State and Materials
  Science}\ }\textbf {\bibinfo {volume} {25}},\ \bibinfo {pages} {100923}
  (\bibinfo {year} {2021})}\BibitemShut {NoStop}%
\bibitem [{\citenamefont {Panda}(2009)}]{panda2009environmental}%
  \BibitemOpen
  \bibfield  {author} {\bibinfo {author} {\bibfnamefont {P.}~\bibnamefont
  {Panda}},\ }\bibfield  {title} {\bibinfo {title} {Environmental friendly
  lead-free piezoelectric materials},\ }\href@noop {} {\bibfield  {journal}
  {\bibinfo  {journal} {Journal of materials science}\ }\textbf {\bibinfo
  {volume} {44}},\ \bibinfo {pages} {5049} (\bibinfo {year}
  {2009})}\BibitemShut {NoStop}%
\end{thebibliography}%
\end{document}